# A Perspective-Based Understanding of Project Success

Laurie McLeod[1], Bill Doolin[2] and Stephen G. MacDonell[1]

[1]*SERL, School of Computing & Mathematical Sciences*
[2]*Business School*

*Auckland University of Technology*
*Auckland, New Zealand*
lauriemcleod@xtra.co.nz, bill.doolin@aut.ac.nz, stephen.macdonell@aut.ac.nz

**Abstract**

*Answering the call for alternative approaches to researching project management, we explore the evaluation of project success from a subjectivist perspective. An in-depth, longitudinal case study of information systems development in a large manufacturing company was used to investigate how various project stakeholders subjectively perceived the project outcome and what evaluation criteria they drew on in doing so. A conceptual framework is developed for understanding and analyzing evaluations of project success, both formal and informal. The framework highlights how different stakeholder perspectives influence the perceived outcome(s) of a project, and how project evaluations may differ between stakeholders and across time.*

**Keywords** - project evaluation; project success criteria; stakeholder perspectives; longitudinal case study; outsourced information systems development

## I. INTRODUCTION

As a general concept, project success has received considerable attention within the project management research literature over the last three decades (Ika, 2009; Pinto & Slevin, 1988). As our understanding of project success has evolved and matured (Jugdev & Müller, 2005), we have come to recognize the complexity and ambiguity that surrounds it, both in terms of its definition and its measurement (Baccarini, 1999; Fowler & Walsh, 1999; Hyväri, 2006; Ika, 2009; Jugdev & Müller, 2005; Thomas & Fernandez, 2008). Project success is now regarded as a multidimensional construct, with interrelated technical, economic, behavioral, business, and strategic dimensions (Bannerman, 2008; Cao & Hoffman, 2011; Ika, 2009; Jugdev & Müller, 2005; Jugdev, Thomas, & Delisle, 2001; Shenhar, Dvir, Levy, & Maltz, 2001; Thomas & Fernandez, 2008), although this is not always evident in the application of project success measurement.

Prior empirical studies of project success have generally been quantitative, often involving surveys comprising exploratory or evaluative questions with Likert scales, analyzed using statistical techniques, and in which success is measured using a simple formula that is unequivocal and easy to apply (Ika, 2009). These studies are grounded in an objectivist tradition, in which it is assumed that a universal, objective set of success criteria exists in practice (Ika, 2009), to be measured using scientific method and quantitative techniques (Fitzgerald & Howcroft, 1998).

Several authors have expressed concern over the predominance of objectivist project management research (Fincham, 2002; Ika, 2009; Packendorff, 1995; Söderlund, 2004). In particular, Cicmil and Hodgson (2006) argue that, despite the level of research attention that project success has received, the characterization of project outcomes as successes or failures remains problematic and, in general, the development of project management knowledge has been fragmented and incomplete. Moreover, attempts by the project management research community to remedy any knowledge shortcomings continue to rely almost exclusively on objectivist approaches (Cicmil & Hodgson, 2006), even if they encompass a more complex or contingent view of projects and project outcomes (Fincham, 2002; Ika, 2009).

These authors call for the emergence of alternative theoretical and methodological approaches to studying project management in order to create "new possibilities for thinking about, researching, and developing our understanding of the field as practiced" (Cicmil & Hodgson, 2006, p. 111; see also Alderman & Ivory, 2011; Fincham, 2002; Ika, 2009; Packendorff, 1995; Söderlund, 2004). In particular, Ika (2009) suggests that a subjectivist research approach offers a viable alternative avenue of research to the objectivist tradition dominating project management research. From a subjectivist viewpoint, project success and failure are social phenomena, subjectively and intersubjectively constructed by individuals and groups of individuals (Alderman & Ivory, 2011; Ika, 2009; Packendorff, 1995). A focus on sense-making and the interpretation of meaning encourages the use of idiographic approaches and qualitative techniques (Fincham, 2002; Fitzgerald & Howcroft, 1998; Ika, 2009). A subjectivist approach to project management research is consistent with calls for more empirical studies (Cicmil & Hodgson, 2006), especially in-depth, longitudinal case studies that address the dynamic nature and social context of projects and project management (Packendorff, 1995; Söderlund, 2004). Such studies can give us deeper knowledge of how different stakeholders perceive projects and their outcomes.

In this article, we address these calls relating to project management research by providing an empirical analysis of a contemporary information systems (IS) project from a subjectivist perspective. Following Ika (2009), we are interested in understanding how the various project stakeholders subjectively perceived project outcomes and the evaluation criteria they drew on in doing so. In undertaking this research, our motivation is to challenge the way researchers and practitioners view project outcomes and how they evaluate projects. A broad and inclusive understanding of success is also important for senior management and project managers, the latter of whom "are constantly trying to define and manage project success in both subjective and objective ways" (Jugdev & Müller, 2005, p. 19), in order to answer questions about their project's progress.

We focus on an IS project as a research setting for two reasons. First IS projects have become a significant and frequently studied setting in the project management literature (Ika, 2009; Kloppenborg & Opfer, 2002; Rivard & Dupré, 2009; Urli & Urli, 2000). Second, contemporary IS development projects frequently involve the active participation of a wide range of internal and external stakeholders (each having their own motivations and goals, and thus potentially their own criteria for success), and the views of some of these groups have not necessarily been considered in prior research (Cicmil & Hodgson, 2006; Haried & Ramamurthy, 2009).

The remainder of the article is structured as follows. In the next section, we consider current thinking on evaluating project outcomes within the project management literature in general, drawing on prior research relating to the IS domain when appropriate. We then describe the case study and methods used in data collection and analysis. Next, we present our empirical findings, before discussing the results of our analysis. The article concludes with the implications of our study for research and practice.

## II. LITERATURE REVIEW

In the project management literature, the outcome of a project is frequently conceived of in terms of "success" or "failure"— although identifying just what constitutes these can be problematic. In general, there is a lack of consensus on how to define success, lack of success, and failure, and despite their frequent use, such terms are perceived to be vague and difficult to measure (Baccarini, 1999; Fowler & Walsh, 1999; Hyväri, 2006; Ika, 2009; Jugdev & Müller, 2005; Thomas & Fernandez, 2008). Further, success (or failure) is not an absolute or "black and white" concept (Wateridge, 1998). Projects may be viewed as successful to varying degrees, depending on which success criteria are met (Baccarini, 1999; de Wit, 1988). In this section, we map out some of the elements that contribute to the conceptual and definitional ambiguity of project success. These include issues of multidimensionality, scope, temporality, perspective, and context.

There have been various attempts over the history of project management to define suitable criteria against which to define and measure project success. Perhaps the most well recognized of these is the long established and widely used "iron triangle" of time, cost, and quality (Atkinson, 1999; Cooke-Davies, 2002; de Wit, 1988; Ika, 2009; Jugdev & Müller, 2005; Jugdev et al., 2001). Although the definition of quality is potentially very broad (Ika, 2009) in relation to the iron triangle, it is often restricted to meeting scope or functional and technical specifications (Agarwal & Rathod, 2006; Baccarini, 1999; Bannerman, 2008; Ika, 2009; Wateridge, 1998). Expressed by the mnemonic "on time, within budget and to specification" (Turner, 1993, p. 76), these criteria constitute economic and technical dimensions of project success. They are popular, particularly in the engineering, construction, and information technology (IT) fields (Ika, 2009) because they can be made objective, tangible, and measurable (Baccarini, 1999; Ika, 2009); they fall within the ambit of the project organization (Pinto & Slevin, 1988); they are short-term, ending upon project delivery (Baccarini, 1999; Ika, 2009; Pinto & Slevin, 1988); and they can be used to evaluate a project manager's performance (Jugdev & Müller, 2005; Wateridge, 1998).

However, as a number of commentators have pointed out, the iron triangle dimensions are inherently limited in scope (Atkinson, 1999; Ika, 2009; Wateridge, 1998). Indeed, a project that satisfies these criteria may still be considered a failure; conversely, a project that does not satisfy them may be considered successful (Baccarini, 1999; de Wit, 1988; Ika, 2009). In particular, the iron triangle has been criticized for its exclusive focus on the project management process and for not incorporating the views and objectives of all (both internal and external) stakeholders (Atkinson, 1999; Baccarini, 1999; Bannerman, 2008; de Wit, 1988; Jugdev & Müller, 2005; Wateridge, 1998). Even if the focus is on the manner in which the project was conducted, several authors have suggested that meeting time, cost, and quality specifications are not the only relevant criteria; for example, project management efficiency and effective project team functioning are also important (Baccarini, 1999; Shenhar & Dvir, 2007; Toor & Ogunlana, 2010).

Within the project management literature, researchers have progressively widened the scope and constituency of what is meant by project success, recognizing that project success is more than project management success, and that it needs to be measured against the overall objectives of the project (Baccarini, 1999; Cooke-Davies, 2002; Ika, 2009; Jugdev & Müller, 2005). This reflects a distinction between the success of a project's process and that of its product (Baccarini, 1999; Markus & Mao, 2004; Wateridge, 1998). Focusing on the latter may lead to consideration of criteria such as product use, user or client satisfaction, and benefits to users or clients (Baccarini, 1999; Bannerman, 2008; DeLone & McLean, 2003; Shenhar, Dvir, Levy, & Maltz, 2001; Wateridge, 1998).

Bannerman (2008) also suggests that within individual project disciplines there are discipline-specific processes and practices to which success measures may be attributed, such as project governance, risk management, change management, and quality management.

Project success has also been extended to encompass the achievement of a broader set of organizational objectives, involving benefits to a wider range of stakeholders, including senior managers and project sponsors (Baccarini, 1999; Ika, 2009; Jugdev & Müller, 2005). In particular, attention has focused on the immediate and direct impact of the project on the organization, including whether the business case and objectives for the project investment have been met and benefits to the business realized (Bannerman, 2008; DeLone & McLean, 2003; Markus & Mao, 2004; Shenhar et al., 2001). Project success may even be extended further to include the accomplishment of more strategic objectives and benefits, including impacts on markets and competitors, business development or expansion, and ability to react to future opportunities or challenges (Bannerman, 2008; Jugdev & Müller, 2005; Norrie & Walker, 2004; Shenhar et al., 2001; Toor & Ogunlana, 2010). For example, Cooke-Davies (2002) and Ika (2009) discuss measuring the success of organizational portfolios or programs of projects that are aligned with corporate strategy.

Project success criteria that focus beyond the project management process constitute behavioral, business, and strategic dimensions. While such criteria can support a more holistic and inclusive definition of success (Jugdev & Müller, 2005), they tend to be subjective, intangible, and difficult to measure (Baccarini, 1999; Ika, 2009; Jugdev & Müller, 2005). Such criteria are particularly important in the IS/IT domain, where considerable emphasis is placed on subjective issues, such as user satisfaction (DeLone & McLean, 2003; Jugdev & Müller, 2005; Petter, DeLone, & McLean, 2008; Wateridge, 1998). Table 1 summarizes the various criteria proposed as the scope of what constitutes project success has expanded. While project success is a multidimensional construct, in practice not all criteria may be considered relevant on all projects, and different criteria may be emphasized on different types of projects (Bannerman, 2008; Shenhar et al., 2001; Wateridge, 1998).

Temporality is an element of the evaluation of project success in two ways. First, the different foci of consideration discussed earlier are meaningful in relation to organizational processes and practices that operate over different time frames. Thus, project management success and its criteria are relevant over the time frame of a project; once a product or service is delivered, product success and its criteria become increasingly relevant as the product or service is used (or not) within its operational environment; while business success and strategic benefits and their criteria are relevant over even longer time frames (Baccarini, 1999; Bannerman, 2008; Jugdev & Müller, 2005; Pinto & Slevin, 1988; Shenhar et al., 2001). Second, individual or collective opinions and evaluative assessments of a project are not necessarily static and may change over time as situations evolve and contexts change (Lanzara, 1999). A project that initially may be deemed a success can subsequently come to be regarded as a failure, or vice versa (de Wit, 1988; Ika, 2009; Wilson & Howcroft, 2002). In recognition of the temporality of project success, various authors suggest that multiple evaluations should be undertaken at different points in time for different purposes (Atkinson, 1999; Jugdev & Müller, 2005; Karlsen, Andersen, Birkel, & Odegard, 2005; Khang & Moe, 2008; Pinto & Slevin, 1988; Wateridge, 1998).

| Expanding dimensions | | |
|---|---|---|
| **Process Success** | **Product Success** | **Organizational Success** |
| *Focus on project management* | *Focus on project objectives* | *Focus on organizational objectives* |
| • On time | • Product use | • Business benefits |
| • Within budget | • Client satisfaction | • Strategic benefits |
| • To scope/specifications | • Client benefits | |

Table 1: Project success criteria.

A number of authors have highlighted the subjective and perceptual nature of project evaluations, and that a person's perception of an outcome is dependent on his or her perspective or viewpoint (Agarwal & Rathod, 2006; Baccarini, 1999; Bannerman, 2008; Fowler & Walsh, 1999; Ika, 2009; Shenhar et al., 2001; Wateridge, 1998; Wilson & Howcroft, 2002). Given that a project has a range of stakeholder groups, each with a particular viewpoint (Baccarini, 1999), individuals or groups are likely to differ in their assessments of the extent to which a project is successful. Taken to an extreme, a success for one group may be perceived as a failure by others (de Wit, 1988; Riley & Smith, 1997). In an IS project, the stakeholders may include senior management, organizational project management office staff, the project manager, the project governance group, the project sponsor, the project owner, project team members, developers, and various user stakeholder groups. While many IS project stakeholders are internal to the organization commissioning the project, some may be external—for example, external consultants, vendors, or developers in an outsourced context (McLeod & MacDonell, 2011).

Each stakeholder group will have its own view of project success, judging it according to different criteria (Agarwal &

Rathod, 2006; Baccarini, 1999; Riley & Smith, 1997; Wateridge, 1998). For example, in an IS project, the project manager and project governance team may focus on the success of the project process, while users are likely to concentrate on the operation and implementation of a project product, considering success in relation to the impact of the IS on their work and organizational roles. Project sponsors may be concerned with the survival of their project (Wilson & Howcroft, 2002) or the activity it was intended to support, while senior management may be interested in the achievement of business objectives and the strategic benefits delivered by the project (Bannerman, 2008; Wateridge, 1998). Technical staff such as developers on the project may view success in terms of product quality and functionality or the opportunities for new skill acquisition and learning that can be carried forward to future projects (Agarwal & Rathod, 2006; Linberg, 1999; Riley & Smith, 1997). External contractors may be concerned with containing project costs and duration (Bryde & Robinson, 2005), and securing further work.

The criteria used to evaluate project success are based on stakeholders' particular expectations of the project (Jiang, Chen, & Klein, 2002; Lim & Mohamed, 1999), with success reflecting the extent to which these expectations are perceived to have been met. In turn, expectations derive from and express value-based beliefs and desires about how a project will serve stakeholders' interests and/or needs whether role-related or personal (Baccarini, 1999; Bannerman, 2008; Lyytinen, 1988; Lyytinen & Hirschheim, 1987). Thus, the assessment of project success is a value judgment (Seddon, Staples, Patnayakuni, & Bowtell, 1999), and "success" becomes a subjective evaluation from a person's perspective rather than an objective description based on independent criteria (Mitev, 2005). Different values, interests, needs, and expectations become relevant to particular stakeholders' interpretations depending on the social, economic, historical, and organizational context in which a project is situated (Bartis & Mitev, 2008; Lyytinen & Hirschheim, 1987).

When confronted with a complex problem such as evaluating a project outcome, individuals engage in sense making, seeking, and interpreting information in order to construct meaning in relation to the project (Galliers & Swan, 2000). This meaning is constantly shaped in response to new knowledge, changing contextual elements, the behavior of others, and an individual's engagement with the project and its product (Constantinides & Barrett, 2006; Giddens, 1984; Walsham, 1993). The individual and shared stocks of knowledge and systems of meaning that individuals use to help them interpret and make sense of a project are based on their past experiences and participation in social processes and professional groups (Kjaergaard, Kautz, & Nielsen, 2007; Luckmann, 2008; Walsham, 1993). Thus, an individual's evaluation of a project outcome may be influenced by (possibly competing) organizational commitments, sectional interests, or professional affiliations (Butler, 2003). Further, any development of a shared understanding of the project outcome necessarily involves the communication and negotiation of individual and collective perceptions, expectations, and evaluations.

This makes the formal evaluation of a project outcome a negotiated, and often political, process. Rather than being discrete, objective outcomes, success and failure are constructed as contested subjective interpretations that may be modified in response to political maneuvering, persuasion, or changes in the organizational and technological context (Bartis & Mitev, 2008; Fincham, 2002; Mitev, 2000; Wilson & Howcroft, 2002). Any apparent definitional closure surrounding a particular project outcome does not necessarily represent consensus or shared interests and values, as not all interpretations or viewpoints may be afforded equal status (Bartis & Mitev, 2008; Riley & Smith, 1997; Walsham, 1993; Wilson & Howcroft, 2002). While some expectations (e.g., client or user expectations) are expressed as project goals and requirements, other expectations may remain unarticulated or only vaguely expressed. The latter may result from the unclear nature of an expectation, the sheer number and diversity of stakeholders involved, or an inability or lack of opportunity for them to voice their expectations (Lyytinen & Hirschheim, 1987). As Thomas and Fernández (2008, p. 733) note, "how success is defined and who evaluates success therefore affects the final judgment of success and failure." Formal evaluations are doubly political, in that the "official narrative" may be used to confer status or stigma, legitimize particular behaviors or courses of action, justify large or risky projects, or distance the future from the past (Bartis & Mitev, 2008; Cicmil & Hodgson, 2006; Fincham, 2002). However, irrespective of the formal evaluation of a project outcome, other informal evaluations that various individuals and stakeholder groups make along the way, based on their specific interests and expectations, influence their decisions and actions, and thus the unfolding project (Walsham, 1993).

In summary, labeling a project outcome as a "success" or "failure" is convenient but overly simplistic. A degree of conceptual and definitional ambiguity surrounds project success. Further, evaluations of project success are necessarily perceptual and (inter)subjectively constructed. This suggests that a subjectivist approach to studying project outcomes would be fruitful; it provides an understanding of how the wide range of project stakeholders in contemporary IS development make sense of a project in relation to their various perspectives, interests, and expectations. The project case study that follows draws on extensive empirical evidence to illustrate this.

## III. THE CASE STUDY AND RESEARCH METHODS

Our objectives in this study were to understand how project outcomes are subjectively perceived by different stakeholders and the evaluation criteria different stakeholders draw on in doing so. To address these objectives, we conducted an in-depth, longitudinal case study of an IS development project in AlphaCo (a pseudonym), a large manufacturing company. The project involved commercial package acquisition, outsourced (but largely on-site) solution development, and a range of internal and external participants and stakeholders. As such, it provides a useful exemplar of development in a contemporary IS environment involving client-vendor relations (Haried & Ramamurthy, 2009).

The project studied involved developing a dynamic, flexible, and scalable database solution to replace existing financial spreadsheet models used to manage AlphaCo's IT outsourcing contract. These models were used for contract evaluation and performance reporting. Figure 1 shows the main participants involved in the project. The project owners were the unit responsible for managing this contract—the IS Outsourcing Management (ISOM) team. This was a small team of business analysts, including Claire (who had developed the original spreadsheet models) and Gary (the intended main user of the new database solution), led by Dave, the ISOM manager, all of whom participated as members of the AlphaCo project team. The project sponsor was James, the IS senior manager to whom the ISOM team reported.

Consistent with organizational policy, an external project manager, Frank, was hired to manage the project through its expected completion by the end of 2005, and external consultants, SoftCo, were engaged to supply a multidimensional database and OLAP tool (MDS, SoftCo's proprietary application development tool) and to develop the desired database solution. SoftCo's main project team was composed of their project manager, Marie, and two senior developers, Nancy and Ross.

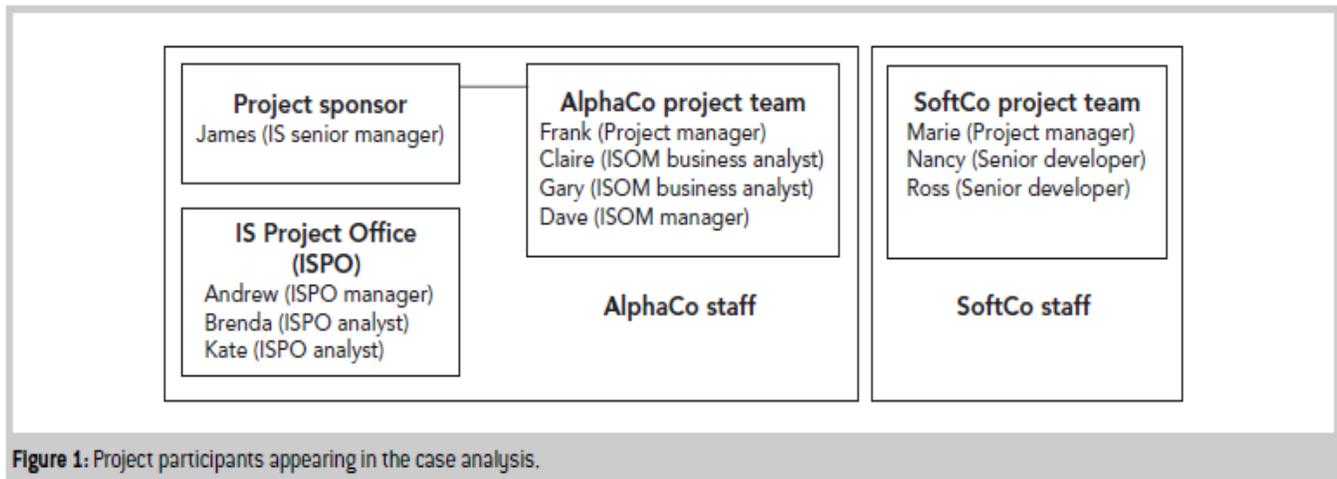

Figure 1: Project participants appearing in the case analysis.

To promote standardization in project management practice, AlphaCo had implemented organization-wide project management standards that required projects to follow a life cycle with defined phases, decision points (gates), and requisite documentation. Tracking, a purpose-built project management support tool that additionally served as a repository for project reporting documents, was also used. For IS projects, these formal project management processes, including an IS project life cycle, were administered by the IS Project Office (ISPO), including the manager, Andrew, and two analysts, Brenda and Kate. The IS project life cycle comprised six project phases (Concept, Feasibility, Planning, Development, Implementation, and Completion), each ending with a decision gate and specific project documents.

The ISOM project was followed through this project life cycle, and into subsequent deployment and use of the MDS solution, from mid-2005 to mid-2007, by the first author. Fieldwork involved an intensive eight-month period of participant observation coinciding with the main project activity. This was followed by a number of site visits as work on the project became more sporadic, until some form of closure was achieved in terms of the use of the new solution. In total, 558 hours were spent on site, observing project activities and meetings, and conducting 33 semistructured interviews with internal and external staff. Interviewees comprised the project participants from AlphaCo and SoftCo mentioned earlier and shown in Figure 1, together with a further five AlphaCo IS managers and analysts. Interviewees were asked a set of general questions about their organizational role, their involvement in the project, their perspective on events and activities, and their experiences of development practice. In addition, a list of specific questions was compiled for each participant about aspects of the project that they could comment on. A number of the interviewees were interviewed multiple times over the course of the project, as new developments or issues emerged. Interviews were typically recorded (with permission) and transcribed in full. Where taping was not possible, extensive field notes were made. All project documentation (including e-mail) was made available to the researcher. Various internal organizational documents and publicly available articles on AlphaCo and its IS function were also reviewed to provide contextual information.

Conducting a longitudinal case study enabled observation of the IS project in its organizational context, the description of events as they occurred, and access to participants' actions and interpretations at the time. A single case study allowed a holistic, in-depth analysis of one setting. Intensive and prolonged interaction with project participants and stakeholders provided insights into local knowledge and practices, and revealed alternative or shifting interpretations of the project outcome (Lanzara, 1999; Nandhakumar & Jones, 1997). Consistent with a subjectivist research approach, the focus was on the participants' experiences and interpretations of the project, and how they made sense of the development process and outcome, including the solution itself, through their behavior and language (Orlikowski & Baroudi, 1991; Putnam, 1983). In doing so, multiple perspectives from participants at different levels in the project and organizational hierarchies were accessed on multiple occasions (Cicmil & Hodgson, 2006).

A comprehensive thematic analysis (Braun & Clarke, 2006) was conducted on the data collected from field notes, the transcripts of interviews and meetings, and project and organizational documents. Data were read and reread multiple times by the first author, categorized and compared across common themes that emerged around aspects of project content, context, and process (Walsham, 1993), and informed by relevant IS development literature. Chronological order and temporal relativity of the data were maintained throughout the data analysis. This facilitated the identification and description of key events and activities within the project, which provided the structure for a detailed "chronological" (Allison & Merali, 2007; Yin, 2003) case narrative of the project as it unfolded over time. Further details of the data collection and analysis methods used in this study can be found in McLeod, MacDonell, and Doolin (2011).

Given the objectives of this article, we focus on the perceptions of different project participants and stakeholders regarding the project outcome, rather than the project itself.

## IV. THE ISOM PROJECT

The ISOM project was intended to be a straightforward migration of existing spreadsheet models to a database solution. From the outset, the project was perceived to be relatively small and well defined, with no major threats to project delivery within the desired project time frame of the end of 2005. This perception was evident in the project's feasibility report, which did not anticipate any problems with the technical development of the project or in obtaining required resources. Management support was "clearly demonstrable," and the users were "committed" and "highly positive." The implementation was considered to be "very simple," and no issues were expected to arise due to the project's size or complexity (Project feasibility report, September 2005). Thus, many of the commonly reported risks associated with project failure did not appear to apply in this case. The only project risk or constraint anticipated in the feasibility report was the availability of Claire and Gary, as key users of the existing spreadsheet models, to interact with the external developers and test the emerging solution. In practice, however, the project was subject to various delays and problems so that it stretched over a protracted period of time, eventually losing much of its currency as organizational events overtook it. In the following, we focus primarily on the completion of the project and the early use of the MDS solution as most relevant to our consideration of project outcomes (Pinto & Slevin, 1988). This is where stakeholder perceptions and opinions about the project outcome tend to stabilize, although they may shift as situations and context change.

### A. Solution Development

Work on the MDS solution began in early November 2005, on site at AlphaCo on a standalone server supplied by SoftCo. It involved input from Frank, the external project manager, and Gary, the intended primary user of the MDS solution. Solution development entailed iterative cycles of building by SoftCo's senior developers, Nancy and Ross, testing by Frank and Gary, and subsequent amendment by the developers. Solution delivery had been negotiated for early December, but development quickly fell behind schedule and milestones had to be revised. By the end of 2005, the expected project completion time and date of Frank's departure (at the end of his contract), the MDS solution was largely complete but untested and lacking documentation.

At the beginning of 2006, Gary took over as the AlphaCo project manager, responsible for getting the MDS solution tested and operational within AlphaCo's networked environment. Reconciliation of the emerging MDS solution with historical data was delayed due to data corruption, problems with the MDS solution, and errors in the original spreadsheet models. In addition, during this time, a major restructuring of AlphaCo IS disrupted standard operations. While the contract performance reporting functionality of the MDS solution was operational from the end of February 2006, the contract evaluation functionality was not completed until mid-May 2006.

### B. Project Closure Under Pressure

During March 2006, Gary came under increasing pressure from various parties to complete the ISOM project. Claire summarized the frustrations and increasing urgency to have an operational MDS solution for use by the ISOM team:

> We've got to get the damn thing working. It's really frustrating. . . . It has to be working. We've got to get it working. So, we've got to get it done this week. . . . The problem has been keeping focused. . . . Busy, busy, busy. (Claire, informal conversation, March 2006)

The ISOM manager, Dave, believed that an operational MDS solution would not only be useful to his team, but would also justify the project and its benefit at an organizational level:

> This is like a long sunset, isn't it? . . . I've put the pressure on Gary to get that thing finished. . . . We're getting things that we'd be able to use it for. . . . We've got to demonstrate that we have finished, we're using it, and the benefits are being realized. (Dave, informal conversation, March 2006)

In addition, the ISPO was pushing Gary to close the project at this time by completing a brief closure report in Tracking, AlphaCo's project management support tool. Information from this report is required to fulfill portfolio reporting to the CIO and his management team. Although the formal Development and Implementation phases of the ISOM project had not been completed (the project's status in Tracking was still Development, to which it had been set in December

2005), institutional pressures were applied to formally close the project with respect to organizational systems, in order to capitalize the project within the current financial year and avoid it registering as a "red light" on the AlphaCo IS balanced scorecard (Gary, informal conversation, April 2006). During February, Gary had revised the forecast project completion date to February 17, 2006, and subsequently recorded the actual project completion date as February 28, 2006 (Project document, March 2006). The day after Gary completed the closure report on April 3, an ISPO analyst updated the project's status in Tracking to "Complete."

As early as February, Marie, SoftCo's project manager, had "[her] Board on [her] back about the outstanding invoices" (Marie, e-mail, February 2006) yet to be paid by AlphaCo and, as the project increasingly dragged on, she exerted pressure on Gary to pay these and sign off on the project. By April, she was requesting the return of SoftCo's server, on which the MDS solution had been developed, threatening to charge AlphaCo for its continued use—a move Gary interpreted as encouraging him to sign off on the project (Gary, interview, May 2006). By mid-May, Gary regarded the MDS solution as "essentially complete." Apart from some reconciliation problems originating from AlphaCo's original models and data, the MDS solution functioned correctly, and Gary was keen to arrange a final meeting with SoftCo:

> The project, in my mind, is essentially complete. . . . I'm going to get a buy-in from the powers that be that it can't be reconciled. . . . Accept it, move on. So, as far as I'm concerned, it's happening as soon as I have some sign-off from Claire, Dave, and a few others. I want to tick it off and ring SoftCo and have a final sign-off meeting. (Gary, interview, May 2006)

This meeting occurred in early June. Discussion focused on organizing a final training session for potential solution users and transferring the MDS solution to AlphaCo's networked environment. For the AlphaCo project team, the meeting represented some sort of closure in terms of the ISOM project: "Fabulous. I think we're cooked. Awesome. Thank you very much. Long time coming, but we're there. We're there" (Claire, project meeting, June 2006). A final full-day training session run by SoftCo was held soon after. From SoftCo management's perspective, this represented a final handover of the MDS solution: "That [training] should be part of the closure. . . . It's very important just to make sure things get closed, rather than just the guys finishing and going 'It's done. See you later'" (SoftCo representative, informal project conversation, January 2006).

Claire and Gary saw the training session as an opportunity to raise awareness of the MDS solution with a wider audience in AlphaCo IS, including other team members, other users of information produced by the MDS solution, and potential users of other MDS applications. In the end, however, only Gary and Vince, another ISOM team member, attended: "You can lead your horse to water, but . . . only Vince went. It would've been far better if Claire and [a non-ISOM analyst], perhaps Dave, had gone along. But everyone always gets pushed for time" (Gary, interview, October 2006). Ironically, Vince was transferred overseas in September 2006 and was subsequently made redundant. Installation of the MDS solution in AlphaCo's networked environment took four months to complete (instead of the usual two weeks), mainly because of delays in the IT infrastructure support team's processing of the necessary change request and server instability problems.

According to AlphaCo IS project management documentation, the ISPO was supposed to instigate a formal review once the project was complete. From the time the ISOM project status was shown as "Complete" in Tracking (April 2006), Gary expected a review to occur (although he was unsure of its form, which would depend on the size, subject, scope, or level of complexity of the project) and was waiting for the ISPO to organize it. However, as part of the AlphaCo IS restructuring occurring at the time, the ISPO merged with another AlphaCo project office, with an expanded company-wide scope. In the consequent "upheaval and change" (Gary, interview, June 2007), a final review was never undertaken:

> Because we had a restructure, we lost that—a whole lot of post-implementation review and all that stuff that was going to happen. . . Someone may put their hand up and say, "Hey, look, we should've done this," but I don't think so. There's bigger priorities. (Gary, interview, March 2007)

This was not an isolated occurrence, as James, the project's sponsor, indicated that the restructuring had disrupted several IS projects (James, interview, June 2006).

### C. (Non)Use of the MDS Solution

The intended users of the MDS solution were the ISOM team, who were responsible for monthly contract performance reporting and ad hoc contract evaluation and scenario analysis. Outside the ISOM team, the main user of information produced from the new solution would be James, the IS senior manager with overall responsibility for vendor, contract, and commercial management. Other IS senior managers, AlphaCo senior management, and the Board, and occasionally individual business units, would be indirect users of information from the MDS solution in a variety of reports, including the AlphaCo IS balanced scorecard.

The restructuring of AlphaCo IS in early 2006 changed organizational reporting requirements, including for the ISOM team, which was reformed into a new (smaller) team with wider responsibilities for financial management and performance reporting across all of AlphaCo IS (rather than their original focus on the IT outsourcing contract) and still reporting to James. However, in March 2006, James was transferred to another organizational unit and was replaced by a new IS senior manager, Stuart. With James's departure, interest outside the original ISOM team in the monthly contract performance report ceased:

> No one has really been viewing it [the performance report], because there's been such disruption outside. But I'm going to have to start getting out there and say, "Look. Who owns this report? Who wants this report?" and "Let's use it." I

mean, it's a good report. It shows us a lot of things. (Gary, interview, April 2006)

Gary continued to upload data into the MDS solution and to produce monthly contract performance reports until mid-2006, although these were not published for external consumption. From June 2006, Stuart initiated a new type of monthly reporting based around cost centers across AlphaCo IS (consistent with a new organizational focus on cost management). In this new reporting climate, the outsourcing contract performance report and even the IS balanced scorecard lost much of their currency: "We got Stuart, who would've got all our reporting, and he wasn't interested in that reporting" (Gary, interview, March 2007). In effect, the restructuring of AlphaCo IS had removed an important source of organizational legitimacy for reporting on the outsourcing contract: "Previously, we were the ISOM team and we reported the outsourced agreement. But there's no ISOM team anymore. . . . You can't produce a report for a team that doesn't exist" (Gary, interview, October 2006). Nevertheless, Gary believed that the functionality provided by the MDS solution would still be beneficial to the company when the disruption caused by the restructuring settled down:

> We've now got to work out, going forward, what reports we need and who needs to see them . . . In about two months, someone is going to say, "Hey, how's [the IT outsourcing provider] tracking against what we expected it to track?" . . . That's where it will be useful. (Gary, interview, October 2006)

However, from the time of the completion of the MDS solution in May 2006 until July 2007 (the end of the fieldwork period), the MDS solution was used only sporadically, as a data repository and to retrieve data for ad hoc tasks: "The MDS solution hasn't been used at all really. I use it for the odd times to spit out some volumes and things like that. It's a good repository, but we could've done that in a [simpler] database" (Gary, interview, March 2007). The full potential of the MDS solution was not realized, in part, because of Gary's incomplete understanding of the contract evaluation functionality of the MDS solution: "It's just limited by my knowledge of how it works" (Gary, interview, March 2007). Further, other uses of the MDS solution, which could have been achieved by expanding its functionality, and the intended extension of the user base of the MDS solution, had not eventuated by mid-2007, partly because of a reduction in staff numbers and the increased demands on those remaining:

> There's lots of things that it could be used for. . . . We're just not very proactive at the moment. . . . It will be useful, but I don't think we've got the amount of staff to actually start doing the proactive stuff. (Gary, interview, March 2007)

Dave suggested that the relatively small size of the ISOM project meant that the MDS solution "fell off the radar when the restructuring occurred" (Dave, informal conversation, March 2007), while larger projects, particularly enterprise-wide ones, had sufficient staff to manage them through the restructuring. By March 2007, Dave felt that the relevance of the outsourcing contract reporting from the MDS solution would once again be realized: "With all the changes, even the demand for some of the reporting has gone away. It went away for a year. . . . But now it's come back" (Dave, interview, March 2007). On Dave's suggestion, once again, Gary began uploading monthly contract performance data files into the MDS solution, which had not been updated since July 2006. Part of this renewed urgency on Dave's part seems to have stemmed from pressure applied by Stuart: "to actually show that it's [the MDS solution] actually useful" (Gary, interview, March 2007). Certainly, Dave continued to feel the need to demonstrate the benefit of the MDS solution, given the investment made by AlphaCo.

For some time, Gary had wanted to demonstrate the capabilities and potential of the MDS solution to others in AlphaCo IS, particularly Dave and Stuart. Gary believed that a major reason the MDS solution was being underutilized was its lack of visibility in the company. He also wanted to review the future use of the MDS solution in the restructured AlphaCo IS context. In June 2007, a meeting was held to demonstrate the MDS solution and discuss its future. Those present at the meeting, including Dave and another IS manager, saw its potential and felt that there would be demand for its use going forward. Crucially, however, Stuart cancelled his attendance at the demonstration meeting, which Gary interpreted as reflecting "the low priority of the [MDS solution]" (Gary, informal conversation, June 2007). This effectively meant that AlphaCo's IS senior management remained unaware of the MDS solution: "Stuart didn't come, and suddenly now you can't take it up any higher, because you've got a glass ceiling. . . . It's just sort of gone in the too-hard basket" (Gary, interview, July 2007).

## V. PROJECT EVALUATION IN ALPHACO IS

There was a perception within AlphaCo that, historically, many IS project outcomes had been problematic. For example, in mid-2004, an internal AlphaCo document highlighted a poor record of project performance in terms of process and product (e.g., "Projects often late, over budget. . . . Systems may not support business needs"). Despite this implicit evaluation of IS project outcomes and that various IS process documents referred to project success, at the time this study was conducted there was not a single, formal definition of project success or failure used within AlphaCo, a limitation recognized by the ISPO manager: "It's a weak area for us, quite frankly" (Andrew, interview, May 2006).

In practice, different success criteria were used for different purposes. For example, in tracking the performance of active projects on a monthly basis—in effect, project management success—the ISPO utilized the traditional iron triangle criteria: "on time, in scope, on budget . . . are quite heavily looked at" (Kate, interview, July 2005). These measures were reported in a monthly project portfolio dashboard to the CIO and his management team, as well as in the AlphaCo IS balanced scorecard. Further, as part of their weekly review of the dashboard, the ISPO used these three criteria to identify potentially problematic projects and to evaluate "whether we need to follow anybody up" (Kate, interview, July 2005).

In terms of overall project success, the criteria seemed to be somewhat broader and less well defined: "We are still arguing about the criteria for [project success]" (Brenda, interview, May 2006). For example, IS process documents referred to process and product quality and the delivery and realization of benefits, in addition to the traditional iron triangle success criteria. Business (client) acceptance of a solution was also emphasized with, for example, AlphaCo's CIO noting that successful development was understood generally as whether the project "meets the criteria as outlined in the business case." There was also some recognition of the possibility of multiple perspectives on project success, in that part of the formal business requirements management process for large or complex projects specified identifying success criteria for each of a project's stakeholders. A formal business benefits realization process was starting to be introduced into AlphaCo IS after mid-2006 (by which time the ISOM database solution had been delivered): "That's coming, but, you know, to be frank, it's something that hasn't been institutionalized yet" (Andrew, interview, May 2006).

In summary, at the time of this study, a range of project success criteria was in use within AlphaCo IS for different purposes and to varying extents. In particular, the ISPO used traditional process success criteria to monitor and evaluate project management performance, while the concept of client acceptance was the most visible product success criterion observed. Evaluation of projects against organizational objectives such as the realization of business benefits was recognized but not implemented.

## VI. STAKEHOLDER PERCEPTIONS OF THE ISOM PROJECT OUTCOME

Although, as noted earlier, a formal evaluation of the ISOM project was never conducted, various project stakeholders made informal evaluations based on their specific perspectives, interests, and expectations. A stakeholder analysis of project outcome evaluation is presented and summarized in Table 2.

### A. AlphaCo ISPO

The ISPO's role was focused on project management processes and reporting. This involved monitoring the progress of IS projects and "maintain[ing] a certain level of consistency in the project delivery" (Brenda, interview, May 2006). As such, the ISPO's concerns related to ensuring compliance to organizational processes "in terms of reporting, in terms of financial control and tracking, in terms of quality assurance. So, really focused on time and budget" (Andrew, interview, May 2006). This emphasis on resource usage and reporting prompted the pressure they placed on Gary to formally close the ISOM project in AlphaCo's project management tracking system. For the ISPO, project success was being able to record the project's status as "Complete," thus avoiding its registering any "danger signs up on our [project portfolio] dashboard" (Brenda, interview, May 2006).

In terms of the iron triangle criteria used by the ISPO, the ISOM project was officially completed to specifications and within budget but ran slightly over time. In his official closure report in Tracking (April 2006), Gary noted that "All objectives . . . have been met. . . . The finishing deadlines for the project were stretched out longer than expected." Indeed, the ISOM project was regarded by the ISPO as more successful than other projects: "There was nothing in the reporting process that we had that suggested I needed to get involved in terms of a project health check" (Andrew, interview, May 2006). Despite the official closure of the project, work on it continued to be performed by Gary and SoftCo in an effort to reconcile the new MDS solution with the original spreadsheet models. One consequence of the official project closure was that the project accounts began to be closed. This meant that costs associated with the additional work, as well as subsequent costs required to transfer the MDS solution to AlphaCo's networked environment, were treated as supplementary. As such, these costs required separate approval; although, had these been included as project costs, the project would still have been within budget.

### B. SoftCo

In terms of project success, the SoftCo developers, Nancy and Ross, were pleased with what they had achieved by the end of 2005: "From my perspective . . . I think Nancy and I worked quite well to get [the MDS solution] done in that time frame" (Ross, interview, December 2005), although they expressed disappointment that it "didn't get delivered on time" (Nancy, interview, December 2005). As developers, their interests focused on the timely delivery of a workable solution to the prescribed specifications. Indeed, the SoftCo development team felt that (barring some minor problems that needed to be addressed) the MDS solution had achieved what it was meant to: "They [SoftCo] do all talk highly of it . . . they are quite positive about it. They believe in it" (Gary, interview, January 2006).

Marie similarly emphasized what had been achieved by the developers, although she pointed out that a constrained project time frame had compromised SoftCo's ability to transfer knowledge of the MDS tool itself: "There was not enough time to get [the ISOM team] up to speed with MDS, to understand how MDS works" (Marie, interview, December 2005). This time constraint was a result of the original contract negotiations undertaken by a SoftCo representative. Influenced by a desire to secure AlphaCo as a major client, the representative underestimated the project's complexity and committed SoftCo to an unrealistic time frame and project budget: "That was a Board decision . . . because obviously AlphaCo's a good client. . . . Usually I'd stick to my guns on time frame, if I don't agree that I can deliver it on time" (Marie, interview, December 2005). As the SoftCo project manager, the iron triangle criteria of on time, within budget, and to specifications were important practical considerations to be met for Marie. She was confident that the developed solution would achieve what it was meant to and, in terms of time and budget, she had always had a realistic assessment of what was achievable: "I always said we could never do it in this time for the money" (Marie, project meeting, December 2005). In terms of her company's objective in establishing a relationship with AlphaCo (for which they were prepared to bear a financial loss on the project), Marie was optimistic: "I think the client relationship was good. . . . You want to keep the relationship going. And I'm hoping that they see that we can deliver" (Marie, interview, December 2005).

| Stakeholder | Perspective/Expectations | Evaluation Criteria | Overall Evaluation |
|---|---|---|---|
| AlphaCo ISPO | Monitoring and reporting project progress and compliance to project management processes | • On time<br>• Within budget<br>• To specification | The project ran without the need for ISPO intervention and was officially closed by its owners within the appropriate financial year |
| SoftCo project team—developers | Timely delivery of a quality product | • On time<br>• To specification<br>• High-quality design | While not meeting the original time frame was disappointing, the developers were satisfied that a high-quality product was delivered |
| SoftCo project team—project manager | Delivering a functional solution to specifications while minimizing cost and time overruns, and establishing a client relationship with AlphaCo | • On time<br>• Within budget<br>• To specification<br>• Client satisfaction | While time and cost overruns were unavoidable, the solution fulfilled stated objectives and was accepted by the client; a good client relationship was established |
| AlphaCo project team—project manager | Effective project management and delivery of a quality solution that achieved objectives | • On time<br>• Within budget<br>• To specification<br>• User satisfaction<br>• Business benefits | Departed before complete solution delivery but anticipated that the eventual MDS solution would fulfill objectives and even exceed client expectations |
| AlphaCo project team—ISOM analysts | A smooth project with timely receipt of a functioning solution that met their needs and enabled them to better perform their roles | • On time<br>• Within budget<br>• To specification<br>• Functioning of the project team<br>• Solution use<br>• User satisfaction<br>• User benefits | The MDS solution took longer to finalize and implement than expected but met their expectations |
| AlphaCo project team—ISOM manager | A functioning solution, in use, with benefits to his team, and a demonstrated business case | • On time<br>• Within budget<br>• To specification<br>• Solution use<br>• User benefits<br>• Business benefits | A useful solution was delivered, but the continued inability to demonstrate its use and benefits was frustrating |
| AlphaCo IS senior management—project sponsor | Successful completion of the project with a usable solution with benefits for the ISOM team | • On time<br>• Within budget<br>• To specification<br>• Solution use<br>• User benefits<br>• Business benefits<br>• Strategic benefits | The MDS solution met functional requirements and represented an improvement over the previous user tool; supported an organizational objective but of limited direct business value |
| AlphaCo IS senior management—IS senior manager | Inherited the project from his predecessor; any solution would need to be relevant to new IS organizational priorities | • Business benefits?<br>• Strategic benefits? | Unknown, but the MDS solution appeared to be of low priority and relevance |

Table 2: Stakeholder analysis of project outcome evaluation.

## C. AlphaCo Project Team

As AlphaCo's external project manager, Frank was "accountable to the project sponsor for the success of the project in terms of time, quality, costs and achievement of objectives" (Project manager terms of reference, April 2005). Part of that role involved defining appropriate criteria for client acceptance and acquiring sign-off of the MDS solution. Accordingly, Frank created a quality assurance sign-off sheet for Claire and Gary (representing the users of the new solution) to assess the MDS solution against nine criteria, including delivery of all the functional specifications and their satisfaction with the solution, training, documentation, and support. The criteria also included perceived business benefits, such as the scalability of the software for other modeling solutions and the transfer of knowledge about the MDS software tool itself. The delays to the project meant that Frank's contract ended before the MDS solution was finalized and implemented. Gary, who assumed the project manager role, was aware of the need to complete the quality assurance sign-off documentation, but this was never done. Throughout the remainder of the project, his main focus was a pragmatic concern to test and implement the MDS solution. Although Frank departed before the project was completed, he believed that SoftCo would "come up with the solution we wanted at the end of the day" (Frank, interview, December 2005), and that MDS generally had the potential for other business applications.

As the main intended users of the new MDS solution, Claire and Gary considered the project to have been successful in terms of both process criteria (although delayed in terms of delivery) and product criteria. For example, when asked about the project, Claire replied: "It's been a good project . . . a bit disjointed at times. . . . Things took a bit longer to complete" (Claire, interview, June 2006). Functioning of the project team and vendor relations were obviously a consideration in her evaluation of the project process, as she added: "We've had good team members. SoftCo have [sic] been a good supplier to work with" (Claire, interview, June 2006). In terms of product success, in interviews and project conversations at various points in the project, Claire and Gary expressed their satisfaction with the MDS solution and their intention to use it. For example, as the solution began to emerge, Claire enthused: "It's fantastic. And there's a hundred and thousand different uses that we can have with this now" (Claire, project meeting, December 2005). After the solution had been completed, she commented: "It's going to be most valuable" (Claire, interview, June 2006). Similarly, in his official project closure report, Gary noted that the MDS solution was "far superior to present solution . . . much improved . . . more accurate and useful. . . . The final result is a very useful application that has endless opportunities" (Project closure report, April 2006). Later, this favorable assessment was repeated with more informal evaluative comments, such as "It's a very usable tool" (Gary, informal conversation, April 2006) and "It's definitely going to add to what we can do in Excel" (Gary, interview, October 2006). As can be seen, many of the evaluative statements that Claire and Gary made about the MDS solution were future-oriented and often focused on the benefits to them as users.

By mid-2006, Gary was a confident user of the MDS solution, at one point noting: "I've got a reasonable knowledge of this thing" (Gary, project meeting, June 2006). However, Gary's subsequent lack of use of the MDS solution meant that he lost much of the detailed knowledge he had acquired: "I don't even understand it [the contract evaluation functionality]. I don't know how it works" (Gary, interview, March 2007). This seems to have adversely affected his assessment of the now "overcomplicated" MDS solution: "Really, we should've been building a model that you just, anyone could utilize without having a heap of training. . . . [It's] not very intuitive" (Gary, interview, March 2007).

Dave, the ISOM manager, also considered the project to be successful. His evaluative criteria included the benefits for the MDS solution's users: "It's something that we need to do to make . . . my team's job easier, to enable us to provide a higher level and more informative level of reporting and information out to the business" (Dave, product demonstration, September 2005). For Dave, the eventual MDS solution meant that the ISOM team could now "stop relying on dumb spreadsheets . . . [and instead] rely on a single, intelligent repository. . . . The reports that we come up with . . . should be more reliable, more complete, and more informative" (Dave, interview, May 2006). Dave defined project process success in terms of the development of a solution that met requirements, thus excluding the problematic and much-delayed testing and deployment of the MDS solution:

> The project was a success. . . . In the project, the solution was built, it was delivered. . . . The final deployment I see as something being quite different, because the solution operates as intended. . . . It worked very well as a project. (Dave, interview, May 2006)

As noted earlier, as the ISOM manager, Dave was also very conscious of the need to demonstrate to others in the organization that the MDS solution was being used and that the business case and objectives for the project investment had been met: "We need to be able to demonstrate to the Finance people that . . . we spent the money and it's finished, and here's the benefit we're getting" (Dave, informal conversation, March 2006). The failure to demonstrate the business benefits of the MDS solution continued to be a concern for Dave: "We need to show something for it [as] on any significant spend" (Dave, interview, March 2007). Interestingly, Gary pointed out a difference between what the MDS solution could do and the sort of demonstrable use of it that Dave envisaged: "I think people have sort of got different expectations. . . . I think that's a big disconnect between everyone" (Gary, interview, March 2007).

## D. AlphaCo IS Senior Management

As the senior manager to whom the ISOM team reported, James was the business owner of the project and its sponsor. The project was funded from his organizational budget, so the traditional success criteria of time, cost, and meeting specifications were relevant from his perspective, although not a day-to-day concern: "[James] just wants to see a result at the end but he's left it up to us to do it. . . . If the budget and things started going over, he would be far more involved" (Gary, interview, January 2006). When asked to comment on whether the project was successful, James noted that the MDS solution met all the requirements and "will be a significant improvement in that area," although "from a delivery [perspective], it's taken longer to implement than initially per the project plan. . . . It seems to have been successful" (James, interview, June 2006). For James, actual use of the solution by the ISOM team would be "one of the real true tests of success" (James, interview, June 2006).

In terms of business benefits, the ISOM project had been considered to be "necessary to maintain AlphaCo's business" and aimed at "extending and defending our core business" (Project document, March 2005). The large size and value of AlphaCo's IT outsourcing contract meant that the project was treated "as a 'must do' exercise" (ISPO document, October 2005), of "critical" priority (Project document, March 2005). However, the changing organizational context over the (delayed) course of the project removed much of that perceived criticality. As Gary commented: "It was critical a year ago, but obviously it doesn't seem to be so critical now" (Gary, interview, October 2006). From James's perspective, the business and strategic benefits of the ISOM project were limited: "It's reasonably remote from the business strategy. . . . You're implementing a tool that's kind of part of a support function, when there's [sic] other projects that . . . are actually deriving direct business value" (James, interview, June 2006).

James's departure to another organizational role left the ISOM project without a champion within IS senior management. His replacement, Stuart, had different priorities and "a completely different perception of what we should be doing" (Gary, interview, March 2007).

In the changed organizational context, which resulted from the Alpha Co IS restructuring, it seems that Stuart did not see the MDS solution as beneficial, something reinforced by his non-attendance at the June 2007 meeting to demonstrate the MDS solution and discuss its future. Without an organizational demand for its use, the business benefits originally envisaged around knowledge transfer and scalability of the MDS software remained unrealized (at least from an external perspective). Further, Gary was the only user with any knowledge of the new MDS solution: "We're back to the same risk. You know, where I'm the only person who knows how to use it" (Gary, interview, October 2006). Tellingly, when Claire needed to model the renegotiation of the IT outsourcing contract, she used Excel spreadsheets, rather than the MDS solution.

## VII. DISCUSSION

The case study provides clear evidence that the evaluation of project outcomes is far more complex than can be conveyed by traditional concepts of success or evaluation criteria— these notions are both limited and limiting. The findings reported here confirm the multidimensional nature of project outcomes and the inadequacy of singular evaluation criteria. They illustrate how project success is evaluated across varying levels of scope in terms of the project's process, product, or organizational impact. The findings illustrate the temporality of project evaluations, with different scopes of evaluation operating over different time frames, and stakeholders' evaluations of project success potentially varying as circumstances change. However, the results of the case study suggest that multidimensionality, scope, and temporality by themselves are insufficient to explain the complexity involved in project outcome evaluation. Critically, evaluating the outcome of a project is a subjective process, in which interpretations of the outcome vary subjectively depending on a stakeholder's perspective and expectations of the project. Such stakeholder expectations are context-dependent. The key findings on project success evaluation from the case study are summarized in Table 3 and discussed below.

The traditional iron triangle criteria of on time, within budget, and to specification were used by almost all project participants at some point in evaluating the success of the project. However, other measures of project process success were also utilized, such as producing a high-quality design and the effective functioning of the project team. Standard project product success criteria, including solution use, user satisfaction, and user benefits, were utilized by members of AlphaCo closest to the project, including the project team (who would be the eventual users of the MDS solution) and the project sponsor. Evaluation of the organizational success of the project—whether immediate or longer term—required the use of broader criteria focused on the achievement of organizational objectives or value, such as business and strategic benefits. As might be expected, the focus and scope of evaluation tended to broaden, moving from the immediate project team staff to higher-level stakeholders, such as project sponsors and senior company managers.

Different types of stakeholder evaluations of project success occurred across different time frames. For instance, while evaluations based on process success criteria were relevant over the project time frame, those utilizing product success criteria required implementation of the new solution. Organizational success criteria became relevant post-implementation, so that the achievement of organizational objectives required a longer time frame for evaluation than that of management of the project process. Moreover, project outcomes were evaluated on an ongoing basis and some stakeholders' assessments of project success changed over time. For example, at the original planned delivery date of the new solution, the project was of concern to the AlphaCo project team—while on budget, the MDS solution had not yet been completed and did not meet specification. Six months later, with acceptance of the MDS solution and sign-off with SoftCo, the project was considered a success by the same stakeholders— although delivered late, it met product success criteria related to a usable and potentially useful solution. However, over the longer term, the inability to demonstrate business benefits arising from the project was frustrating to members of the team such as the ISOM manager.

As Table 2 summarized, the different stakeholders' perspectives shaped their expectations of how the project met their needs or served their interests. For example, the external developers from SoftCo were concerned with the quality, functionality, and timeliness of solution delivery—all process success criteria—while the project managers from both companies had additional interests in client or user satisfaction, reflecting a need to assess the success of the project's product. Similarly, as members of the AlphaCo project team, the ISOM analysts were concerned with process success but as users of the new MDS solution, they were also interested in product success. While some expectations were explicitly formulated as goals and objectives, others were not. Dave, the ISOM manager, perceived a need to justify the company's investment in the project, which meant that his evaluation of the project outcome was shaped by the ability to demonstrate the usefulness and benefits of the resultant solution.

| Project Success Element | Key Findings |
| --- | --- |
| Multidimensionality | • A range of project success criteria were utilized in project evaluations by various stakeholders |
| | • The traditional iron triangle criteria still formed the basis of most stakeholders' evaluations, although other project process success criteria were also utilized |
| | • Standard project product success criteria were primarily utilized by the client organization |
| | • Evaluation of the organizational success of the project required the use of broader criteria, such as business or strategic benefits |
| Scope | • Shifting focus from the project's process to its product and organizational impact reflected an expanding scope of project success evaluation |
| | • Generally, the higher the organizational level of stakeholders, the broader the scope of their project success evaluations |
| Temporality | • As the scope of evaluation expanded, the time frame required for evaluation increased |
| | • Individual and/or collective project success evaluations changed over time in response to changing circumstances and as different scopes of evaluation became relevant |
| Perspective | • Different stakeholders based their evaluations of project success on criteria that matched their expectations of the project |
| | • Different stakeholders' perspectives also influenced their focus of evaluation on process, product, and/or organizational success |
| Context | • The times and places in which the project occurred shaped stakeholders' expectations and evaluations of the project |
| | • Institutional structures, roles, and requirements were particularly influential in shaping stakeholders' expectations in relation to the project |

Table 3: Key findings from the case study.

Further, the project studied did not exist in a vacuum but took place over time and in particular settings composed of different people, practices, and organizational structures, which influenced stakeholders' expectations and perceptions of the project's success. For example, the eventual restructuring of AlphaCo IS reduced the perceived relevance and business value of the new solution. Within this changed organizational context, Gary's experience of using (and not using) the MDS solution led to a shift in his subjective evaluation of the solution. The significance to SoftCo, a relatively small consulting firm, of potentially acquiring the much larger AlphaCo as a client provided a contextual influence not only on SoftCo's actions in securing and conducting the project, but also on their evaluation of its outcome in terms of the ongoing relationship developed with AlphaCo. The ISPO's focus on monitoring the project management process reflected its institutional role and responsibilities within AlphaCo, as well as an organizational history of poor project performance. The pressure exerted on Gary to close the project before delivery of a completed solution reflected the ISPO's expectations around monitoring and compliance reporting; a closed ISOM database project best served the ISPO's needs and the company's formal reporting requirements at that time. Thus, the evaluation of the project's success involved a negotiated and political process that privileged institutional interests over those of the ISOM team.

In turn, stakeholders' expectations influenced their choice of which criteria were perceived to be relevant for evaluating the project's success. Thus, the ISPO's expectations around compliance with project management standards and the achievement of targets or deliverables influenced their use of established and relatively easily measured criteria such as on time and within budget. Other measures of project success, such as client acceptance, were not directly monitored by the ISPO, but left to other stakeholders. Of course, project stakeholders, such as the ISPO, did not necessarily devise the criteria they used in evaluating the project but drew on existing criteria available to them from project management literature and practice, such as the iron triangle.

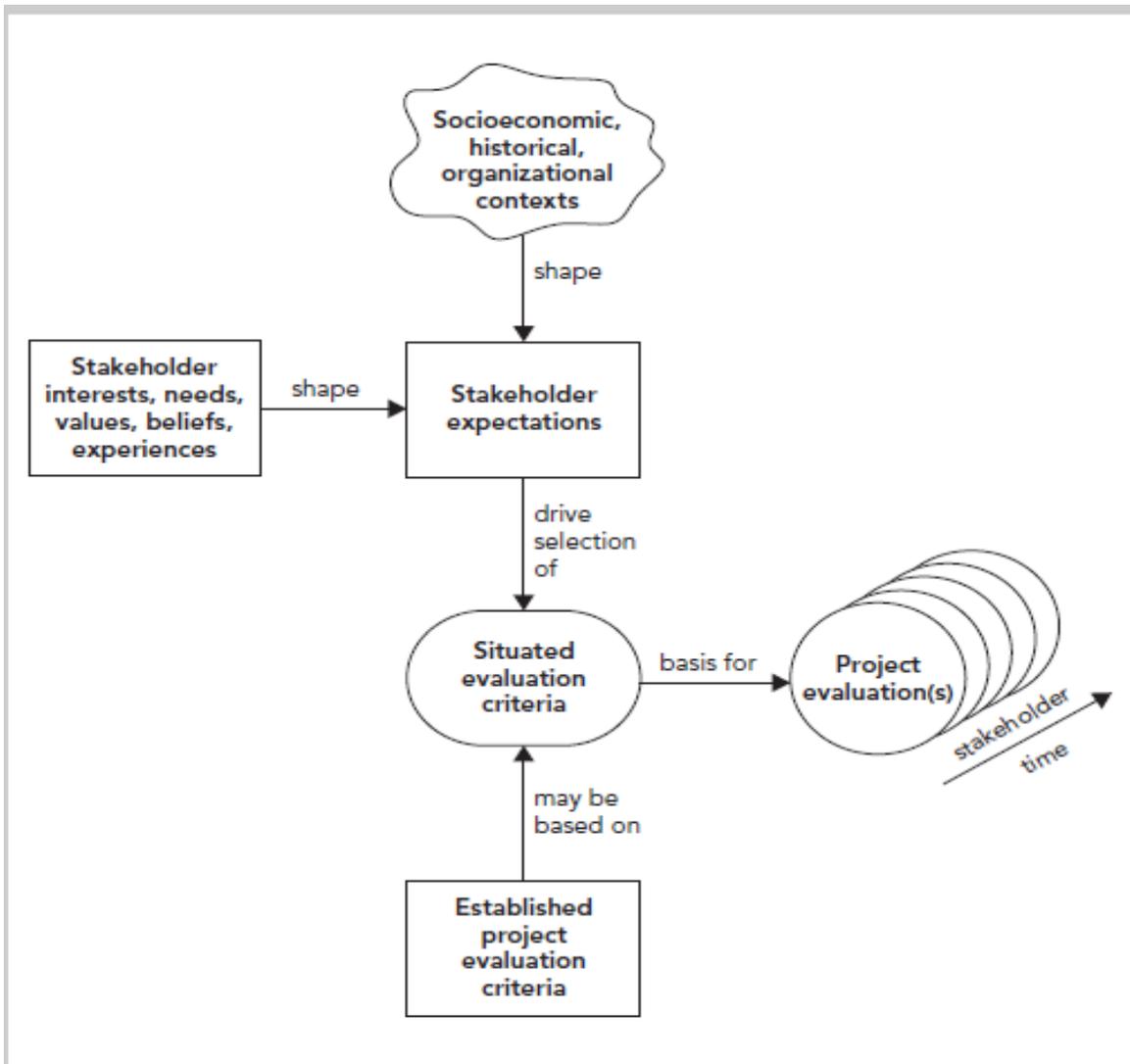

**Figure 2**: A perspective-based framework for evaluating project success.

Overall, there was a degree of convergence around the perceived success of the project across different stakeholders when the scope of the evaluation was focused on the production of a successful solution. Over the time frame of the project, product success was more important to the various stakeholders than the initial failure to achieve the timely delivery of a functioning solution. With the departure of many of the project participants after implementation of the MDS solution the focus of the remaining stakeholders shifted to the business and strategic benefits to the organization. While Gary and Dave remained confident of the organizational value of the MDS solution, the project sponsor and AlphaCo IS senior management appeared less convinced.

Figure 2 presents a conceptual framework that synthesizes our perspective-based understanding of the evaluation of project success. The focus of the framework is on the situated evaluation criteria that form the basis for sense making in relation to project outcome evaluation. Using the term "situated" acknowledges that the form and nature of evaluation criteria are interrelated with, and inseparable from, the material and social circumstances in which they are deployed (Suchman, 2007). That is, while stakeholders may draw on pre-existing, formalized evaluation criteria (cf. Fitzgerald, Russo, & Stolterman, 2002), their selection and appropriation of specific criteria are driven by their expectations of the project and its product. These expectations are shaped by the socioeconomic, historical, and organizational contexts in which the project and stakeholders are located, and by value-based beliefs about how a project will serve stakeholders' perceived interests or needs. The conceptual framework in Figure 2 acknowledges that, rather than a discrete project outcome, multiple project evaluations are possible, which may differ between stakeholders or across time. It follows, then, that instead of talking about a singular project, or even a complex multidimensional one, it is potentially useful to consider a multiplicity of projects, each subjectively perceived and situated in particular contexts and practices. The concept of multiplicity (Mol, 1999) helps to explain why stakeholders can participate in a common activity such as IS development yet understand and evaluate it differently in relation to their particular perspectives and expectations.

## VIII. CONCLUSION

Our aim in this article was to explore the utility of a subjectivist approach to studying project success. Underlying such an approach is the notion that evaluations of project outcomes need to be viewed against individual or collective stakeholder expectations about how a project and its product will serve their needs and interests. From a subjectivist perspective, project evaluation is a complex, ongoing process of sense making, emerging from observations or experiences before, during, and often after a project. Such sense making involves (inter)subjective judgments based on evaluative beliefs and desires, at least some of which become relevant, depending on the context in which the project and its evaluation are situated.

We investigated how project stakeholders subjectively perceived project outcomes through an in-depth longitudinal case study of an IS development project in a large manufacturing company. The case study confirms that project outcomes are interpreted differently from different stakeholder perspectives, and also potentially at different times, and are constructed through subjective processes of sense making. The criteria judged appropriate by a particular stakeholder for evaluating a project are context- and perspective-dependent and may reflect dimensions that variously focus on a project's process, product, and/or organizational objectives. Based on our findings from the case study, we developed a perspective-based framework for understanding the evaluation of project outcomes (Figure 2). The framework focuses attention on the complex process by which project evaluations emerge over time and across multiple stakeholders, who have varying expectations shaped by societal, organizational, and work contexts.

The perspective-based framework for project evaluation has a number of implications. It serves as a useful analytical device for researchers investigating the subjectivity of project evaluation and the role that context plays in this process. It can provide a theoretical basis for explaining how different stakeholders can produce multiple, potentially conflicting evaluations of a project outcome. The notion of multiplicity is particularly powerful, as it offers a conceptual apparatus for understanding that different stakeholders experience and subjectively perceive a particular project in potentially distinct ways. The framework can also be employed as a sensitizing device by a researcher conducting empirical research on project management, providing criteria to guide data collection, analysis, and presentation.

With respect to practice, the framework can be used by project managers and other practitioners to recognize and understand that project evaluation is an emergent, multidimensional, situated, and subjective process. The possibility of multiple perspectives, and hence multiple evaluative judgments, could be usefully applied by project management practitioners in the planning, management, and conduct of formal project evaluations. For example, acceptance of the validity of multiple perspectives encourages the inclusion of others' perspectives in an evaluation and consideration of a wider range of evaluation criteria. A single, client-driven perspective may not capture the meaning and nature of project success for all the stakeholders involved in a project (Haried & Ramamurthy, 2009).

Recognition that individuals and groups may informally evaluate a project at various points across and beyond the life cycle of a project, utilizing a range of possible evaluation criteria, suggests that project managers should seek feedback from stakeholders regularly during a project. This may assist managers to preempt or address emergent problems (e.g., around user-related issues) or avoid escalating commitment to problematic project trajectories. In addition, an awareness of the evaluation of a project's organizational success by high-level stakeholders helps project managers focus not only on the project management process, but also on the broader organizational objectives for the project. Acknowledgment of the contextual elements of a project, such as the organizational history of project experiences and product use, would facilitate the organizational learning from past projects that is a critical part of future project endeavors.

Finally, the research reported in this article has some limitations. It is an initial empirical exploration in response to Ika's (2009) call for an alternative, subjectivist avenue of project management research. Our findings are based on a single IS project case study, although we believe that the insights gained from our analysis are likely to be relevant to understanding the evaluation of other types of projects in other settings. A subjectivist approach to project evaluation involves a commitment to empirical, qualitative, and preferably longitudinal research, entailing time-consuming fieldwork and detailed qualitative data analysis. Accepting the possibility of multiple narratives of project success and failure eschews the convenience of a single narrative depicting the "true" nature of a project's outcome and acknowledges that such narratives privilege particular stakeholders' perspectives and evaluations (Bartis & Mitev, 2008; Fincham, 2002), including those of the researcher. While this makes comparison between multiple projects and studies more difficult, the generation of empirical case studies utilizing a subjectivist approach will build a body of cumulative research that can address the complexity of project success and increase the inclusion of multiple perspectives in project performance and evaluation, leading to more beneficial outcomes for all project stakeholders.

## ACKNOWLEDGMENTS

Our thanks to the case study organizations and the individuals involved. This research was supported by the New Zealand Tertiary Education Commission's Bright Future Scheme. Preparation of this manuscript was supported by an AUT University postdoctoral fellowship.

Yin, R. K. (2003). *Case study research: Design and methods* (3rd ed.). Thousand Oaks, CA: Sage.

**Laurie McLeod** completed her PhD at the Auckland University of Technology in New Zealand. Her doctoral research focused on a process approach for understanding information systems development and acquisition. Her research has been published in various journals, including *ACM Computing Surveys*, the *Australasian Journal of Information Systems, Empirical Software Engineering*, and the *European Journal of Information Systems.*

**Bill Doolin** is a professor of technology and organization at the Auckland University of Technology in New Zealand. His research focuses on the processes that shape the adoption and use of information technologies in organizations. This has involved work on information systems in the public health sector and electronic business applications and strategies. His work has been published in international journals such as *Electronic Markets*, the *European Journal of Information Systems*, the *Information Systems Journal*, and the *Journal of Information Technology*.

**Stephen G. MacDonell** is a professor of software engineering and director of the Software Engineering Research Laboratory (SERL) at the Auckland University of Technology in New Zealand. He was awarded BCom(Hons) and MCom degrees from the University of Otago and a PhD from the University of Cambridge. He undertakes research in software metrics and measurement, project planning, estimation and management, software forensics, and the application of empirical analysis methods to software engineering data sets. He is a member of the IEEE Computer Society and the ACM and serves on the editorial board of *Information and Software Technology*.